\documentclass{aa}
\usepackage{graphicx}
\usepackage{txfonts}
\def\mso{\,\mathrm{M}_\odot}

 \def\kms{\, {\rm km}\, {\rm s}^{-1}}

 \def\simle{\mathrel{\hbox{\rlap{\hbox{\lower4pt\hbox{$\sim$}}}\hbox{$<$}}}}
 \def\simgr{\mathrel{\hbox{\rlap{\hbox{\lower4pt\hbox{$\sim$}}}\hbox{$>$}}}}
 \def\msoy{\, \mso~{\rm yr}^{-1}}

\begin{document}
   \title{Constraints on gamma-ray burst and supernova progenitors through circumstellar absorption lines }
                                                                                                   
   \author{A. J. van Marle
          \inst{1}
          \and
          N. Langer
          \inst{1}
          \and
          G. Garc{\a'i}a-Segura
          \inst{2}
          }

   \offprints{A. J. van Marle}

   \institute{Astronomical Institute, Utrecht University, 
              P.O.Box 80000, 3508 TA, Utrecht, The Netherlands\\
              \email{A.vanMarle@astro.uu.nl}\\
              \email{N.Langer@astro.uu.nl}                           
         \and
             Instituto de Astronom{\a'i}a-UNAM, 
             APDO Postal 877, Ensenada, 22800 Baja California, Mexico\\
             \email{ggs@astrosen.unam.mx}
             }

   \date{Received <date> / Accepted <date>}

   \abstract{
Long gamma-ray bursts are thought to be caused by a subset of exploding Wolf-Rayet stars. 
We argue that the circumstellar absorption lines in early supernova 
and in gamma-ray burst afterglow spectra may allow us to determine the main properties
of the Wolf-Rayet star progenitors which can produce those two events.
To demonstrate this, we first simulate the hydrodynamic evolution of the circumstellar 
medium around a 40 $\mso$  star from the creation and evolution of a wind-blown, 
photo-ionized bubble around the star up to the time of the supernova explosion.
Knowledge of density, temperature, and radial velocity of the circumstellar matter as function 
of space and time allows us to compute the column density in the line of sight to the 
centre of the nebula, as a function of radial velocity, angle, and time.
While without radiative transfer modeling and without detailed knowledge of the spatial 
distribution of chemical elements we cannot produce spectra, our column density 
profiles indicate the possible number, strengths, widths, and 
velocities of absorption line components 
in supernova and gamma-ray burst afterglow spectra.
Our example calculation shows four distinct line features during the Wolf-Rayet stage, 
at about 0, 50, 150-700, and 2200$\kms$, with only those of the lowest and highest 
velocity present at all times.
The 150-700$\kms$ feature decays rapidly as a function of time after the onset of the
Wolf-Rayet stage. It consists of a variable 
number of components, and, especially in its evolved stage, 
depends strongly on the particular line of sight.
A comparison with absorption lines detected in the afterglow of \object{GRB 021004} 
suggests that the high velocity absorption component in \object{GRB 021004}
may be attributed to the free streaming
Wolf-Rayet wind, which is consistent with the steep density drop 
indicated by the afterglow light curve. 
The presence of the intermediate velocity components implies
that the duration of the Wolf-Rayet phase of the progenitor of \object{GRB 021004}
was much smaller than the average Wolf-Rayet life time, 
which strongly constrains its progenitor evolution.

   }

  \titlerunning{Absorption lines in GRB and SN spectra}
  \authorrunning{van Marle et al.}
  \maketitle
%

\section{Introduction}

 Planetary nebulae are the result of wind-wind interaction, with a fast wind emitted by a hot post-AGB star sweeping up a previously emitted slow outflow (e.g., Villaver et al. \cite{VMG02}). 
 Stars above $\simeq 25...30\mso$ also undergo a red-blue evolution in the Hertzsprung-Russell diagram, moving from the red supergiant or Luminous Blue Variable (LBV) stage into the Wolf-Rayet phase (Meynet \& Maeder \cite{MM00}).
 As, in both cases, dense outflows occur first at low and then at high speed, 
 Wolf-Rayet stars are expected to produce circumstellar nebulae just like post-AGB stars
(Garc{\a'i}a-Segura et al. \cite{GML96} and \cite{GLM96}), which are indeed observable in a large fraction of them (Miller \& Chu \cite{MC93}).
                                                                                                   
 Unlike AGB stars, massive stars end their evolution in a violent event triggered by the collapse of the massive iron core in the stellar interior.
 While it cannot be excluded that some of the Wolf-Rayet stars quietly collapse into a black hole, it is believed today that Type~Ib/c supernovae and long gamma-ray bursts are both in fact exploding Wolf-Rayet stars. 
 In Type~Ib/c supernovae,  photospheric hydrogen appears to be absent, while their location and their late time spectra are consistent with core collapse events (van Dyk \cite{D92}; Clocchiatti et al. \cite{CWBF96}; Wang et al. \cite{WBHW03}).
 Also hydrogen-rich Wolf-Rayet stars are potential supernova progenitors (Langer \cite{L91}).

 For gamma-ray bursts, Type~Ic supernovae were found to dominate the late afterglow emission in two cases (Galama et al. \cite{Getal98}; Stanek et al. \cite{SMGetal03}), consistent with the predictions of the collapsar model for long gamma-ray bursts (Woosley \cite{W93};  MacFadyen \& Woosley \cite{MW99}).
 Rigon et al. (\cite{Retal03}) argue, based on possible GRB associations with the peculiar Type~II supernovae \object{SN~1997cy} and \object{SN~1999E}, that some hydrogen-rich supernovae may also be related to gamma-ray bursts.

 In the supernova and the afterglow cases, the explosion creates a powerful source of UV...IR photons, either in the supernova photosphere or in the jet interaction with its surroundings.
 While in gamma-ray bursts the photons produced by the interaction of the explosively ejected material with the circumstellar matter  --- e.g., the gamma-ray burst afterglows --- probe the innermost parts of the wind bubble (Chevalier et al. \cite{CLF04}), it is the absorption of photons in the line of sight to the central light source which traces the structure of the whole bubble and thus carries full information about the evolution of the progenitor star. 
                                                                                                   
 With the optical and near-infrared spectroscopy of the afterglow of \object{GRB 021004} starting only 0.6 days after the burst, it has become possible for the first time to use absorption line systems to analyze the entire circumstellar bubble surrounding a gamma-ray burst, as proposed by Schaefer et al. (\cite{Setal03}), Mirabal et al. (\cite{Metal03}), Fiore et al. (\cite{Fietal05}) and Starling et al. (\cite{Setal05}).
 With continuously refined GRB and afterglow observing techniques, more and better afterglow spectra might provide more such cases soon.

 Since Type~II and Type~Ib/c supernovae also have massive star progenitors, one may also expect narrow circumstellar lines in their spectra. 
 Indeed, Dopita et al. (\cite{DECS84}) attributed narrow P Cygni profiles seen in H$\mathrm{\alpha}$ and H$\mathrm{\delta}$ in the spectrum of the Type~IIL~\object{SN 1984E} to a Wolf-Rayet wind with a velocity of about 3000 \mbox{km} \mbox{s}$^{-1}$.
 Bowen at al. (\cite{BRMB00}) and Fassia et al. (\cite{Fetal01}) note the presence of a broad absorption feature at $\sim$ 350 $\kms$ in the spectrum of the Type~IIL~\object{SN 1998S}, which, as they argue, may be caused by a moving shell.
 Dedicated searches for circumstellar absorption lines in very early supernova spectra might again provide more and better cases.
  
 It is the main aim of this paper to demonstrate that circumstellar lines in supernovae and GRB afterglows can be used not only to identify the Wolf-Rayet nature of their progenitor stars, but also to strongly constrain some of its main properties. 
 Here, we present a 2D-hydrodynamic model of the evolution of the circumstellar medium surrounding a massive star, improving upon the models produced by Garc{\a'i}a-Segura et al. (\cite{GML96}, \cite{GLM96}).
 Using this model, we then calculate the column density of the circumstellar medium as a function of the radial velocity throughout the evolution, which allows a comparison with observed circumstellar spectral lines. 
                                                                                                
 This article is structured as follows:
    In Sect. \ref{sec-evolsim} we give an overview of the method we use for our numerical simulation of the circumstellar medium.
    In Sects. \ref{sec-mstoRSG} and \ref{sec-WR} we present the result of these simulations.
    In Sects. \ref{sec-coldens} and \ref{sec-compo} we explain the method to compute the column density as  function of radial velocity.
    Finally, in Sect. \ref{sec-discus} we discuss the result of our simulations and in Sect. \ref{sec-finrem} we summarize our main conclusions.
                                                                                                   
   \begin{table*}
   \label{tab:windpar}
      \caption{
              Assumed parameters for the three evolutionary stages of our 40 \mbox{M$_\odot$} model (Z=0.20).
              }
      \begin{tabular}{p{0.3\linewidth}lllll}
         \hline
         \noalign{\smallskip}
             Phase & End of phase [\mbox{yr}] & duration [$10^5$\,yr] & $\dot{m}$ [$\msoy$] & $v_{\rm wind}$ [$\kms$] & $n_{\mathrm{photon}}$ [\mbox{s}$^{-1}$] \\
         \noalign{\smallskip}
         \hline
         \noalign{\smallskip}
                                                                                                   
          Main Sequence    & 4.309$\times10^6$ & 43.1 & 9.1$\times10^{-7}$ & 890 & 4.62$\times10^{47}$ \\
          Red supergiant   & 4.508$\times10^6$ & 1.99 & 8.3$\times10^{-5}$ & 15  & 3.00$\times10^{41}$ \\
          Wolf-Rayet       & 4.786$\times10^6$ & 2.78 & 4.1$\times10^{-5}$ & 2160& 3.86$\times10^{47}$ \\
                                                                                                   
         \noalign{\smallskip}
         \hline
      \end{tabular}
   \end{table*}

\section{Numerical technique and assumptions}
\label{sec-evolsim}
    At solar metallicity, stars in the mass range of about 25 $\mso$ to 40 $\mso$ are thought to develop into red supergiants and thereafter to become Wolf-Rayet stars.
    This means that three interactions take place in the circumstellar medium (CSM):
  \begin{enumerate}
    \item An interaction between the fast, low density main-sequence wind and the interstellar medium. \\
    \item The slow, high density Red Super Giant wind then hits the hot bubble created by the main-sequence wind. \\
    \item Finally, the massive, high velocity Wolf-Rayet wind sweeps up the previously created structures. \\
  \end{enumerate}
    For the hydrodynamical simulations, we used the ZEUS 3D code by Stone and Norman (\cite{SN92}).
    We simulated the main sequence and early red supergiant (RSG) phase only in 1D. 
    The Wolf-Rayet wind interaction was computed in 2D, so we took the end result of the 1D simulation and map this onto a 2D grid, as described in Garc{\a'i}a-Segura, Langer \& Mac-Low (\cite{GLM96}). 
    However, since we followed the evolution of the circumstellar medium up to the supernova stage, we calculated the whole circumstellar bubble in 2D, instead of cutting out the inner part.
    For simplicity, we divided the stellar evolution into three stages: main sequence, red supergiant and Wolf-Rayet. 
    As input model we used the 40 $\mso$ star model with metallicity 0.02 as calculated by Schaller et al. (\cite{SSMM92}). 
    The resulting parameters for the stellar wind and photon count are given in Table \ref{tab:windpar}.
     
    Mass-loss rate, wind velocity and the number of ionizing photons were taken as average over these three stages. 
    The mass-loss rate follows directly from the evolutionary model, as does the photon count, for which we used a black body approximation. 
    The wind velocity was chosen so that the kinetic energy output of the star during each stage is reproduced. 
    We chose to use the kinetic energy rather than the momentum, since wind driven shells are energy driven for the largest part of their existence (Weaver et al. \cite{WEA77}).
    The density of the interstellar medium was set at $20^{-22.5}$ \mbox{g} \mbox{cm$^{-3}$}.  
    The effect of photo-ionization was included in the simulation by calculating the Str{\a"o}mgren radius along each radial gridline and correcting the temperature and mean particle weight within this radius as described by Garc{\a'i}a-Segura \& Franco (\cite{GF96}) and Garc{\a'i}a-Segura et al. (\cite{GLRF99}). 
    A similar calculation for a 60 $\mso$ star was recently performed completely in 2D by Freyer et al. (\cite{FHY03}).

   \begin{figure}
   \centering
   \resizebox{\hsize}{!}{\includegraphics[width=0.95\textwidth,angle=-90]{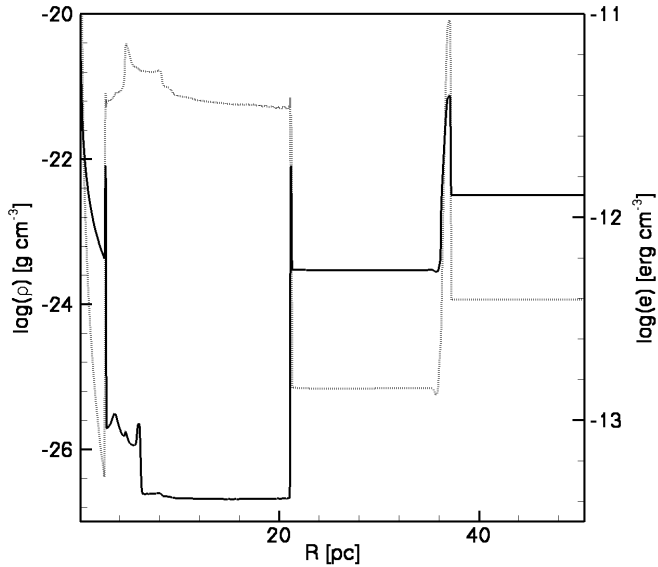}}
      \caption{Structure of the circumstellar bubble around our  40 \mbox{M}$_\odot$ star at the end of the red supergiant phase. 
              This figure gives density (continuous line) and thermal pressure (dotted line) as a function of the radius. 
              From left to right we have: the freely expanding RSG wind, the wind termination shock with the thin RSG shell, the hot bubble, the energy driven shell, the former \ion{H}{II} region, the 'old' shell (which was driven into the interstellar medium during the main sequence phase and is now dissipating, since the photo-ionization no longer extends beyond the wind bubble), and the interstellar medium. 
              }
         \label{fig:RSGbubble}
   \end{figure}

\section{The CSM during main sequence and RSG phase}
    \label{sec-mstoRSG}
    From the beginning of the main sequence, the \ion{H}{II} region created by radiation from a massive star pushes a shell into the interstellar gas. 
    This shell consists of neutral hydrogen and is produced by a D-type front; i.e., the gas in the shell enters the ionization front subsonically.
    At the same time the kinetic energy of the wind is converted into thermal energy through the collision with the surrounding matter. 
    This creates a hot bubble which pushes a shell into the \ion{H}{II} region. 
    Eventually, the \ion{H}{II} region and the wind-blown bubble reach pressure equilibrium. 
    Together they drive a shell into the interstellar medium. 
    The shell between the wind blown bubble and the \ion{H}{II} region disappears, though the density discontinuity remains.
    
    This sequence of events depends on the pressure balance between wind driven bubble and \ion{H}{II} region. 
    If the pressure in the wind-driven bubble becomes so high that it drives the shell into the \ion{H}{II} region with supersonic speed, then the two regions will not find a pressure balance. 
    In that case the wind-driven shell will sweep up the \ion{H}{II} region. 
    The end result will be a single bubble without density discontinuity, driving a shell into the interstellar medium. 
    The density profile will resemble that shown in Garc{\a'i}a-Segura et al. (\cite{GLM96}), although the photo-ionization will cause instabilities in the shells (Garc{\a'i}a-Segura \& Franco \cite{GF96} and Garc{\a'i}a-Segura et al. \cite{GLRF99}). 
    The deciding factors here are the ram pressure of the stellar wind, the number of ionizing photons, and the density of the interstellar medium, which together determine the thermal pressure in the bubble and the radius of the surrounding \ion{H}{II} region. 
    Since the ram pressure of the wind depends on both wind velocity and mass loss rate, which in turn depend on such factors as stellar mass and metallicity, the ram pressure can vary considerably. 
    The same is true for the number of high energy photons and the density of the interstellar medium. 
    These two factors determine the radius of the \ion{H}{II} region. 
    If this radius is large enough, the wind-driven shell will have slowed down to subsonic speeds before it reaches the photo-ionization driven shell.
    In that case pressure balance will be reached after all, and a bubble with internal density discontinuity will be created.
    In this paper we shall continue to use the model with a two-part bubble, to show the widest range of possible phenomena in the circumstellar medium. 
    
    During the red supergiant (RSG) phase, the number of ionizing photons is low, so the \ion{H}{II} region disappears. 
    The wind-blown bubble keeps its high pressure and pushes a new shell into the surrounding medium.
    A third shell forms at the location of the termination shock due to the lower ram pressure and much higher density of the stellar wind.
    At the end of the red supergiant phase, the circumstellar medium is built up as follows (see also Fig: \ref{fig:RSGbubble}): 
    closest to the star is the freely expanding RSG wind, which ends at about 3 \mbox{pc} in the wind termination shock. 
    The shock itself is marked by a thin shell of shocked RSG wind material. 
    Next comes the hot isotropic bubble, which in turn pushes a shell (r $\sim$ 21 \mbox{pc}. in Fig. \ref{fig:RSGbubble}) into the former \ion{H}{II} region. 
    Finally, at about 37 \mbox{pc}, there is the old shell, which was pushed into the interstellar medium during the main sequence phase, but is now slowly dissipating, since it is no longer supported by thermal pressure. 
    The velocities of these three shells is low. 
    In fact, the old shell at 37 \mbox{pc} is standing still, since there is no longer any force to drive it outward. 
    The shell at 21 \mbox{pc} moves outward slowly ($v \simeq 10..20 \kms$), driven by the thermal pressure in the hot bubble. 
    The movement of the shell at the wind termination shock varies through the RSG phase. 
    During the early part of this phase, the wind termination shock moves inward, due to the decreased ram-pressure of the stellar wind. 
    The shell is formed during this phase. 
    The thermal pressure in the hot bubble decreases over time, causing the wind termination shock (and the shell at that location) to move outward with a velocity of about 10 $\kms$.
    For stars with a mass of 25 $\mso$ or less, this is the end of the evolution, as was shown in van Marle, et al. (\cite{MLG04}).

   \begin{figure}
   \centering
   \resizebox{\hsize}{!}{\includegraphics[width=0.95\textwidth,angle=-90]{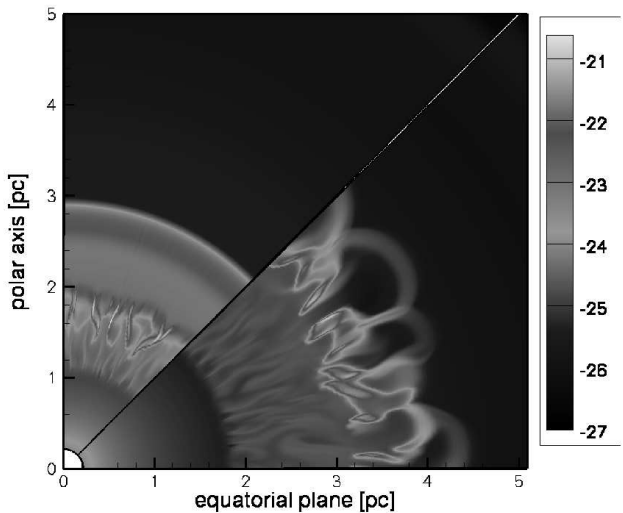}}
      \caption{The logarithm of the density [\mbox{g} \mbox{cm}$^{-1}$] of the circumstellar medium close to the star, just before (upper part, t = 4.514 Myr) and just after (lower part, t = 4.522 Myr) the collision between the RSG shell and the Wolf-Rayet shell. 
       As can be seen, the Wolf-Rayet shell is already fragmented due to Rayleigh-Taylor instabilities before the collision. 
       After the collision, the remnants of the two shells move outward together. 
       }
         \label{fig:coll}
   \end{figure}

   \begin{figure}
   \centering
   \resizebox{\hsize}{!}{\includegraphics[width=0.95\textwidth,angle=-90]{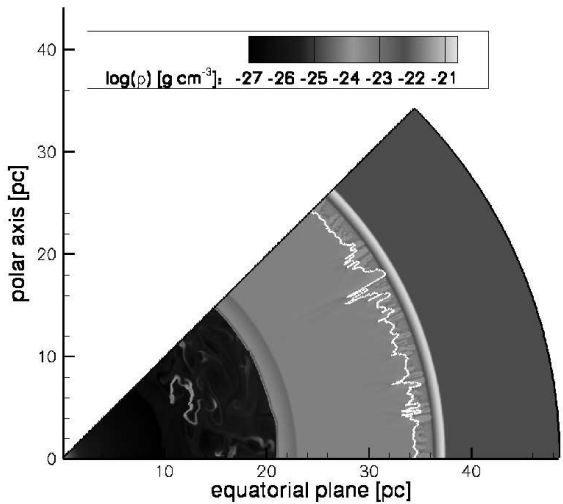}}
      \caption{Circumstellar bubble around a 40 \mbox{M}$_\odot$ star during the second part of the Wolf-Rayet stage (WR2).
      This figure shows the density of the circumstellar medium after 4.566~Myr. 
      Moving outward from the star, we have the freely expanding wind, the wind termination shock, the hot bubble, which drives a shell into the surrounding \ion{H}{II} region, and finally the old shell. 
      The white line shows the Str{\a"o}mgren radius. 
      The blobs of matter inside the hot bubble are the remnants of the RSG shell and Wolf-Rayet shell after collision (cf. Fig. \ref{fig:coll}).
      They are dissipating into the surrounding medium. 
       }
         \label{fig:WRbubble}
   \end{figure}

   \begin{figure}
   \centering
   \resizebox{\hsize}{!}{\includegraphics[width=0.95\textwidth,angle=-90]{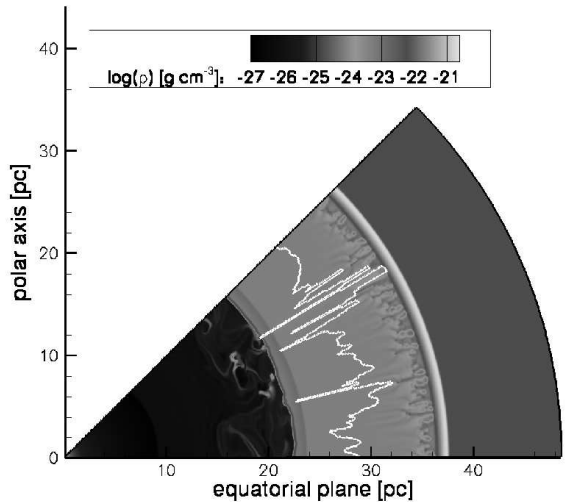}}
      \caption{Similar to Fig. \ref{fig:WRbubble}, but at at 4.598~Myr. 
      The Str{\a"o}mgren radius is receding as the Wolf-Rayet wind adds more matter to the circumstellar medium.
       }
         \label{fig:WRbubble2}
   \end{figure}

   \begin{figure}
   \centering
   \resizebox{\hsize}{!}{\includegraphics[width=0.95\textwidth,angle=-90]{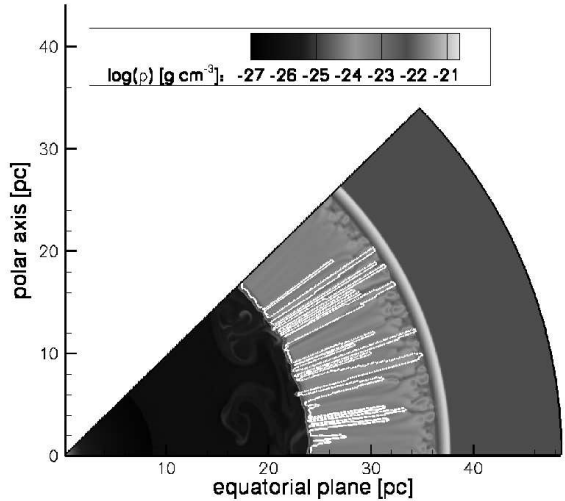}}
      \caption{Similar to Figs. \ref{fig:WRbubble} and \ref{fig:WRbubble2}, but during the third part of the Wolf-Rayet stage (WR3; \mbox{t} = 4.630~Myr).
      The Str{\a"o}mgren radius is now at the same distance as the wind driven shell. 
      The remnants of the RSG shell and Wolf-Rayet shell have almost completely disappeared. 
       }
         \label{fig:WRbubble3}
   \end{figure}

\section{The CSM during the Wolf-Rayet phase}
    \label{sec-WR}
    The period that is of interest to us is the Wolf-Rayet phase, which can be subdivided roughly into three parts, based on the evolution of the circumstellar medium rather than the evolution of the star itself. 
    During the first part (WR1), the ram pressure of the wind rises dramatically as the wind velocity increases (see Table \ref{tab:windpar}).
    This creates a new shell, which moves rapidly through the free-streaming RSG wind and collides with the RSG shell at the wind termination shock (see Sect. \ref{sec-mstoRSG}).    
    Both shells are broken up during the collision, and their shattered remains coast into the surrounding bubble.
    This process was described in great detail by Garc{\a'i}a-Segura et al. (\cite{GLM96}). 
    The collision can be seen in Fig. \ref{fig:coll}. 
    The sudden increase in the number of ionizing photons causes most of the former \ion{H}{II} region (see Fig. \ref{fig:RSGbubble}) to re-ionize.

    The second part of the Wolf-Rayet phase (WR2) is a period of stabilization. 
    The isobarity of the wind-blown bubble was severely disrupted during WR1, and it takes some time to re-establish a constant pressure, since the remnants of the collided shells move into the bubble at velocities close to the speed of sound. 
    During WR2 these remnants travel outward into the bubble, where they eventually dissipate. 
    The high density of the Wolf-Rayet wind makes it difficult for the ionizing photons to penetrate deeply into the circumstellar bubble, causing the Str{\a"o}mgren radius to move inward. 
    The Str{\a"o}mgren radius will eventually end up at the same radius as the shell that is still being pushed outward by the wind-blown bubble (see Fig. \ref{fig:WRbubble} and \ref{fig:WRbubble2}).
    As the kinetic energy of the Wolf-Rayet wind is very high, the thermal energy of the wind-blown bubble will increase, causing the shell that it pushes to accelerate to a velocity of about 50 $\kms$.

    During the third part of the Wolf-Rayet phase (WR3), there is only one hot bubble maintained by both the stellar wind and the photo-ionization. 
    This bubble pushes a shell into the surrounding medium (former \ion{H}{II} region). 
    Photo-ionization causes the shells to fragment (Garc{\a'i}a-Segura et al. \cite{GLRF99}), i.e. the radiation will penetrate through the moving shell at some location and start ionizing the material beyond. 
    The result is an extremely fractured Str{\a"o}mgren radius (see Fig. \ref{fig:WRbubble3}).
    As the pressure in the wind-blown bubble decreases and the moving shell sweeps up more material, it will decelerate. 
    It reaches the original main sequence shell (see Fig. \ref{fig:RSGbubble}) almost at the same time as the central star explodes as a supernova.
    Note that not every Wolf-Rayet star necessarily goes through all three phases. 
    It depends on the total time the star spends as a Wolf-Rayet star, which varies greatly depending on the metallicity and initial mass of the star.

   \begin{figure}
   \centering
   \resizebox{\hsize}{!}{\includegraphics[width=0.95\textwidth,angle=-90]{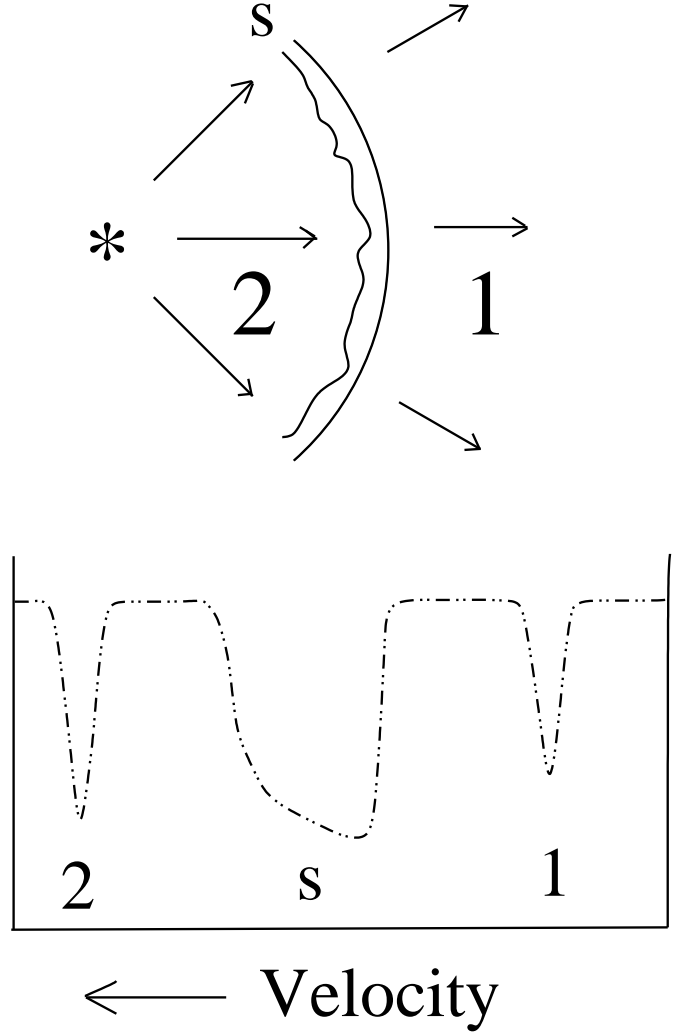}}
      \caption{Schematically, the column density profile produced by a typical wind-wind interaction.
      Here, a slow wind (1) is overtaken by a much faster wind (2). 
      If the collision is supersonic, a shell (S) is created. 
      Winds give narrow absorption lines, since all particles have about the same velocity; only 
      thermal broadening may somewhat increase the width of the line.
      Shells, on the other hand, contain material that is slowing down, so the  material will be spread out over a larger velocity range, creating a much broader line.
      Wind-wind interactions tend to last only a short period of time, but a similar situation occurs when the material in region 1 is stationary.  
      }
         \label{fig:cdcartoon}
   \end{figure}

\section{CSM column densities} 
   \label{sec-coldens}
      A gamma-ray burst afterglow can be described as a point source in the center of our grid. 
      In order to compare the column densities, we take a single radial line in this grid. 
      Moving outward from the center and at each grid point, we take the local radial velocity and the column density (density $\times$ radial length of the grid cell). 
      This gives us the column density as a function of radial velocity, which is equivalent to the absorption as a function of the blue-shift. 
      Typically, the column density as a function of the velocity will look like Fig. \ref{fig:cdcartoon}. 
      Winds give narrow lines, while shells produce a much broader profile.
      Since the circumstellar medium shows considerable variations, we do this for all radial lines in the grid and take the average, in order to make sure that all observable features are accounted for. 
      
      Temperatures in the circumstellar medium can rise as high as $T \simeq 10^8$ \mbox{K}. 
      Therefore, it is necessary to spread the column density given in each grid point over a velocity range, dictated by the Maxwell-Boltzmann distribution for the thermal velocity in a single direction: 
  \begin{equation}
  \label{equ:MB}
      \mathrm{P} (\varv_\mathrm{r}) = \biggl( \frac{m}{2\pi kT} \biggr)^{1/2}e^{-m{\varv_\mathrm{r}}^2/2kT},
  \end{equation}     
      with $m$ the particle mass, $k$ the Boltzmann constant, and $T$ the local temperature. 
      The probability that a particle has a velocity $\varv_\mathrm{r}$ is given by $\mathrm{P}(\varv_\mathrm{r}) \mathrm{d}\varv_\mathrm{r}. $
      In order to calculate the column density as a function of velocity, we first calculate the Maxwell-Boltzmann distribution for the local temperature. 
      Multiplying the distribution function by the mass density gives us the mass density of particles moving within a certain velocity interval. 
      We use a velocity interval $\Delta \varv_\mathrm{r}$ of 1 $\kms$, so the mass of gas particles with thermal velocities in the interval $\Delta \varv_\mathrm{r}$ is:      
  \begin{equation}
  \label{equ:MB_discrete}
     \rho \mathrm{P} (\varv_r) \Delta \varv_\mathrm{r} = \rho \biggl( \frac{m}{2\pi kT} \biggr)^{1/2}e^{-m{\varv_r}^2/2kT} \Delta \varv_\mathrm{r},
  \end{equation}  
      with $\Delta \varv_r$ equal to 1 $\kms$.
      In the co-moving frame, the result is a Gaussian curve with mean velocity zero; i.e., the amount of gas moving away from us equals the amount of gas moving toward us.
      We then shift this entire curve until the mean velocity equals the radial velocity of the gas, as found in the hydrodynamical simulation. 
      Multiplying this by the length of a grid cell ($\mathrm{dr})$ gives us the column density of mass in a velocity interval as a function of the velocity for one grid cell.
      We repeat this process for all the grid cells along a radial line and integrate them for each velocity interval separately.
      The quantity we end up with is $\mathrm{d}_{\mathrm{c}}(\varv,\Delta \varv)$: 
   \begin{equation}
   \label{equ:cdperinterval}
      \mathrm{d}_{\mathrm{c}}(\varv,\Delta \varv) = \int_{\mathrm{r}=0}^{\mathrm{r}=\mathrm{R}} \rho (\mathrm{r}) \mathrm{P}(\varv_\mathrm{r}) \Delta \varv_\mathrm{r} \mathrm{dr},
   \end{equation}   
      with $\mathrm{R}$ the total radial size of the grid.
      This is the column density resulting from all particles along a radial line moving with velocities within 0.5 $\kms$ of $\varv_\mathrm{r}$.

\subsection{Angle-averaged column densities}

 Figure \ref{fig:cd_all} shows the angle-averaged CSM column density as function of the radial velocity over the entire Wolf-Rayet period; i.e., in order to ensure that all features are contributing, we repeat the process described above for all 200 radial grid lines and take the average. 
 To obtain the total column density within a given velocity interval, we need to specify the minimum and maximum velocity and then add up the column densities of the individual 1 $\mathrm{km} \mathrm{s}^{-1}$ intervals between those two.

 The chosen time interval in Fig. \ref{fig:cd_all} starts at the end of the red supergiant phase and ends at the pre-supernova stage of the underlying stellar model.
 While the central star is a red supergiant, there is only one clear line feature, at a radial velocity of nearly zero. 
 There are no high speed features, since the RSG wind, the shells, and the gas in the hot bubble, all move at low velocities. 
 As the Wolf-Rayet wind starts, two new features appear: 
 1)~the Wolf-Rayet wind itself, traveling at about 2200 $\kms$, and
 2)~the region where the Wolf-Rayet wind collides with the much slower RSG wind (i.e. the Wolf-Rayet shell). 

 During the initial Wolf-Rayet phase (WR1 in Sect. \ref{sec-WR}) three absorption features are produced: the zero velocity feature, which was already visible during the RSG phase, a narrow strong line at high velocity (2200 $\kms$) caused by the Wolf-Rayet  wind, and a feature at intermediate velocity, which is first produced by the swept-up red supergiant wind shell and later by its collision with the RSG shell (cf. Fig.~2). 
 Initially, the intermediate velocity feature extends over a large  velocity range (100...600~$\kms$). 
 This is due to both the high temperature (which causes the particle velocities to be spread over a large velocity range) and the fact that gas moving at a velocity of 2200 $\kms$ is slowed down abruptly. 
 However, this feature becomes narrower over time. 
 The bulk material of the shell moves outward at a velocity of about 200 $\kms$ and eventually hits the RSG shell (see Fig. \ref{fig:coll}).
 After the shell collision, the largest fragments maintain their original speed, which is between 150 and 300 $\kms$. 
 In some places shocked Wolf-Rayet wind material breaks through the fragmented shells to form smaller clumps, which can have velocities of up to ca. 700 $\kms$.
 This marks the beginning of phase WR2. 

 During this phase the number of visible features is variable. 
 There is still the intermediate velocity feature caused by the remnants of the RSG and Wolf-Rayet shell. 
 However, since this feature is then caused by shell fragments which cover only a fraction of the possible viewing angles (cf. Figs. \ref{fig:WRbubble2} and \ref{fig:WRbubble3}), it may or may not show up in an observation (see below). 
 As the shell fragments start to dissipate into the surrounding medium, the intermediate velocity feature disappears.

 At about the same time, a new feature appears close to the zero velocity line. 
 This feature, which shows a velocity of about 50 $\kms$, is the result of the wind driven shell which moves outward as the pressure in the wind blown bubble increases (see Sect. \ref{sec-WR}).
 Since this shell is not fractured, the line will always be present, although  it may be difficult to identify due to its proximity to the zero velocity feature.

 The period during which the intermediate velocity feature exists and the 50 $\kms$ feature is formed is shown in Fig. \ref{fig:cd_blowup}, which is a magnified part of Fig. \ref{fig:cd_all}. 
 It shows that the intermediate velocity feature is concentrated between 150 and 300 $\kms$, with subcomponents of up to ca. 700 $\kms$ present during the first 40~000\,yr of the Wolf-Rayet stage. 
 Figure \ref{fig:cd_cuts} shows vertical cuts of Fig. \ref{fig:cd_all} for the times which correspond to the CSM density plots in Figs. \ref{fig:WRbubble} to \ref{fig:WRbubble3}.                                                                                
 During the third WR phase, WR3, the 50 $\kms$ feature disappears as the shell moves further outward, sweeping up material and slowing down.
 In the end only two lines are visible, one at zero velocity and one at the velocity of the Wolf-Rayet wind.

 Any ion which is present throughout the circumstellar medium should show up in a set of absorption features as described here.
 The relative height of the lines, or depth of the absorption, is more difficult to estimate as it depends on the percentage of the particle density of the gas that the ion takes up, which depends both on the chemical composition of the gas and on the degree of ionization (cf. Sect. \ref{sec-GRB021004}). 

\subsection{Line of sight dependence}

\begin{figure*}
   \centering
   \includegraphics[width=0.8\textwidth,angle=-90]{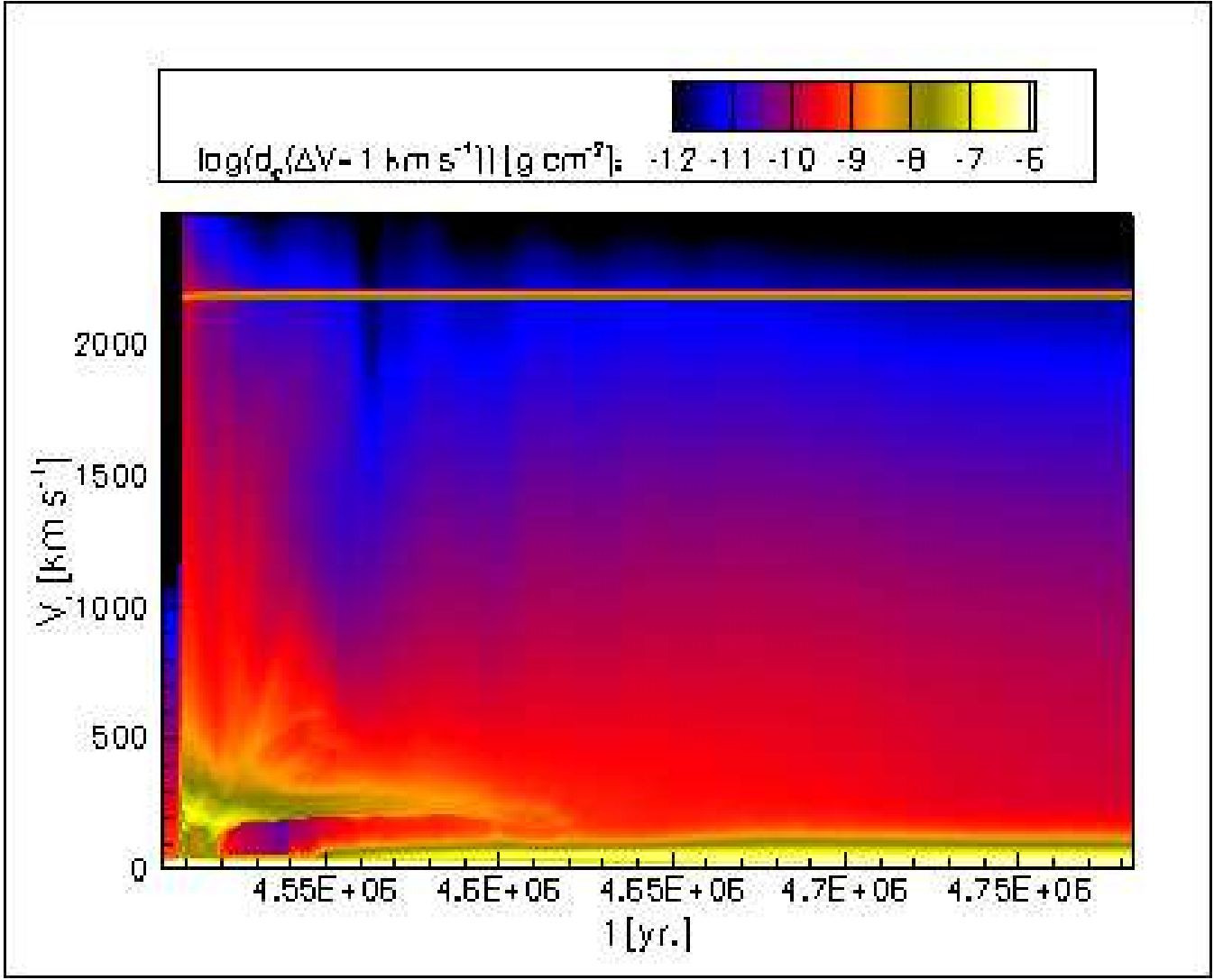}
      \caption{Column density $\mathrm{d}_{\mathrm{c}}(\varv,\Delta \varv)$ with $\Delta \varv$ = 1 $\kms$ of the circumstellar medium around a Wolf-Rayet star as a function of radial velocity and time, averaged over 200 radial grid lines.
      On the horizontal axis: the time since the birth of the star in years. 
      On the vertical axis: the radial velocity in $\kms$.
      The plot starts at the end of the red supergiant phase, when there is only matter moving at low velocity. 
      As the Wolf-Rayet phase sets in a narrow feature at the Wolf-Rayet, wind velocity appears as does a much broader feature at intermediate velocity, caused by the collision between Wolf-Rayet and RSG wind. 
      This feature disappears after about 80\,000 years, when the remnant of the Wolf-Rayet shell dissipates into the circumstellar bubble. 
      The low velocity line splits at about the same time into two separate lines close together, as the increased thermal pressure of the wind bubble drives a shell into the surrounding medium. 
      }
         \label{fig:cd_all}
   \end{figure*}

\begin{figure*}
   \centering
   \includegraphics[width=0.6\textwidth,angle=-90]{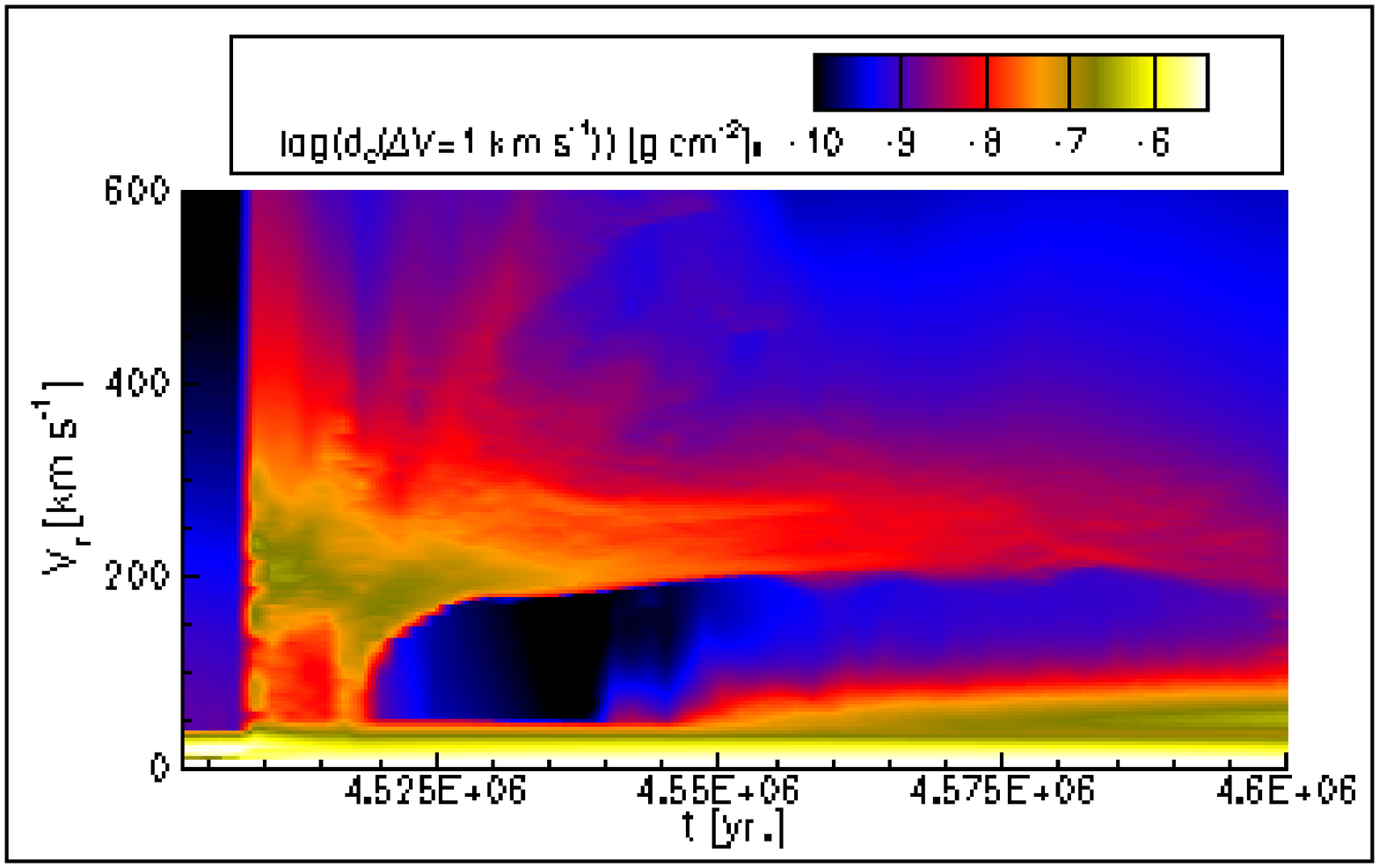}
      \caption{Blow-up of the low velocity part of Fig. \ref{fig:cd_all} for the period during which the intermediate velocity feature is visible. 
      This feature starts at about 4.508 Myr and is the result of the shell swept up by the Wolf-Rayet wind. 
      Initially, the column density between the zero velocity feature and the intermediate velocity feature is quite high, as slow-moving RSG wind material is swept up by the shell and accelerated. 
      After the collision between the Wolf-Rayet shell and the RSG shell (\mbox{t} = 4.52 Myr), this situation no longer applies. 
      The fragments of the shells no longer sweep up a significant amount of mass. 
      Therefore, no matter is accelerated and the column density between the zero velocity feature and the intermediate velocity feature decreases. 
      The intermediate velocity feature is strongest between 150 and 300 $\kms$, but fainter components occur at velocities up to 700 $\kms$.
      At the end of this period, the zero velocity feature splits up into two separate parts. 
      The new feature, showing material moving at about 50 $\kms$, is created by the wind-bubble-driven shell, which moves out into the \ion{H}{II} region. 
      }
         \label{fig:cd_blowup}
   \end{figure*}

 The inclination of the faster parts in the intermediate velocity feature in Figs. \ref{fig:cd_all} and \ref{fig:cd_blowup}, resp. --- i.e. the fact that these features do not appear as vertical structures, but are somewhat inclined --- may lead us to expect multiple absorption components at certain times. 
 However, since the displayed column densities are averaged over all angles, the presence of multiple components of the intermediate velocity feature in the angle-averaged representation does not warrant visibility along a single line of sight (see below).
 On the other hand, even if the angle-averaged representation does not indicate multiple components for a given time, they might well exist at this time for specific lines of sight. 
 We therefore provide a discussion here of the line of sight dependence of the various features.
 The zero velocity line, the Wolf-Rayet wind line, and the line caused by the intact shell (at 50 $\kms$) can be observed independent of the angle. 

 The intermediate velocity feature caused by the fragmented Wolf-Rayet and RSG shells is only visible if a major fragment is in the line of sight. 
 Its probability is strongly dependent on time, as can be seen in Fig. \ref{fig:angledep}. 
 In this figure we have taken the average column density of the gas moving at velocities between 150 and 300 $\kms$, between 400 and 450 $\kms$, and between 450 and 600 $\kms$.
 The first interval contains most of the shell fragments; the last contains the clumps with moderately high velocity (cf. Fig. \ref{fig:cd_blowup}).
 If both these intervals contain a high column density while the intermediate  interval does not, they may show up as separate absorption components in an observed spectrum.

 As Fig. \ref{fig:angledep} shows, the intermediate velocity feature is visible at nearly any angle throughout the entire velocity interval before the Wolf-Rayet shell collides with the RSG shell (t $\simeq$ 4.525 $\times$ 10$^6$ years). 
 After the collision, the shell remnants are broken up into independent, confined fragments. 
 During this period the feature can only be observed if such a fragment happens to be in the line of sight. 

 The probability of such an occurrence is estimated in Fig. \ref{fig:stat}.
 This figure shows the percentage of radial gridlines, along which the column density at 200 $\kms$ is more than five times as high as the column density at 100 $\kms$.       
 As can be seen, the chance of observing the line is highest before the shell collision;  afterwards, it drops rapidly. 
 The increased probability of observing the intermediate velocity feature at about 4.57 million years (Fig. \ref{fig:stat}) is due to dissipation of the shell fragments. 
 These fragments reach the edge of the hot bubble, at which time they are unable to continue their radial movement and start to spread out in an angular direction.  
 However, by then the column density has dropped appreciably, and the velocity tends to zero.

 Figure \ref{fig:angledep} clearly shows the complexity of the intermediate velocity feature in the angle and velocity space. 
 While based on a 2D simulation with limited resolution, quantitative statements are difficult, it appears clearly from this figure that may expect to observe multiple velocity components in the interval of 100 to 700 $\kms$ for certain times and viewing angles. 
 It is obvious that the clumps have considerable angular velocities, which means that they may drift in front of each other during their movements in the hot bubble. 

 Figure \ref{fig:oneangle} shows the column density as a function of the velocity along a single line of sight, at a time shortly after the collision of the
 WR shell and the RSG shell. Time and angle were chosen from inspection of Fig. \ref{fig:angledep}, such that the lower (150...300\,$\kms$) and higher (450...600\,$\kms$) velocity intervals of the intermediate velocity feature are populated, but the middle one (400...450\,$\kms$) is not (pink color in Fig. \ref{fig:angledep}). 
 And indeed, Fig. \ref{fig:oneangle} suggests that an observer looking at the specified time and angle might see two separate components of the intermediate velocity feature.
 However, it is also clear from Fig. \ref{fig:oneangle} that the number of observed separate components may depend sensitively on the spectral resolution of the observation.
 It should be emphasized that compared to Figs.~\ref{fig:cd_all},~\ref{fig:cd_blowup}, and~\ref{fig:cd_cuts}, which show averages over all angles, the structure of the intermediate velocity feature is much more complex when viewed along a single line of sight.

\section{Chemical composition of the CSM}
   \label{sec-compo}
     The column densities calculated in Sect. \ref{sec-coldens} are based on the assumption that a given ion is present throughout the circumstellar medium.
     This, of course, depends on both the chemical composition and the temperature of the material. 
     The composition of the freely expanding Wolf-Rayet wind, which produces the 2200 $\kms$ feature, changes over time as the star evolves from the WN into the WC phase (Maeder \& Meynet \cite{MM94}). 
     The composition of the wind will depend on the initial metallicity of the Wolf-Rayet star. 
     During the WN phase, the wind will consist mostly of helium with some heavier ions. 
     As the star becomes a WC star, carbon (produced by He burning) appears at the surface, becoming the dominant component of the wind.
     
     The wind blown bubble has a mixed composition, since it contains all the material that the star has ejected during its evolution. 
     The remnants of the RSG and Wolf-Rayet shell, which cause the intermediate velocity line, have the composition of the RSG and early Wolf-Rayet phase, plus some material of the hot bubble that they have swept up as they travel outward. 
     The main source of heavy elements here is the red supergiant wind.
     The energy driven shell on the edge of the bubble, the \ion{H}{II} region, and the old main sequence shell, all have the composition of the interstellar medium.
     
     The degree of ionization of the circumstellar gas depends on both temperature and photo-ionization. 
     Ignoring photo-ionization, the temperature of the Wolf-Rayet wind would be comparatively low. 
     The wind leaves the star with a temperature of about 10$^5$ \mbox{K} but would cool down rapidly. 
     Ionization in this area comes from the radiation of the star. 
     The temperature of wind-blown bubbles depends on the kinetic energy of the wind and density of the medium into which the bubble expands. 
     Before the Wolf-Rayet shell collides with the RSG shell, temperatures in the bubble that drives the Wolf-Rayet shell can rise as high as 10$^8$ \mbox{K}. 
     After the collision, the remnants move into the bubble created by the main sequence wind, where temperatures are lower. 
     As the bubble becomes isotropic, the temperatures in the high density areas will be lower, but the temperature of the bubble as a whole increases due to the high kinetic energy of the Wolf-Rayet wind. 
     Throughout the bubble, temperatures between 10$^5$ \mbox{K} and 10$^7$ \mbox{K} can be found. 
     The energy driven shell and the area beyond form an \ion{H}{II} region, which has a temperature of about 10$^4$ \mbox{K}. 
     The old shell and the interstellar medium are cold. 
     This is the situation before the central star collapses. 
     However, we should keep in mind that the emission from the gamma-ray burst afterglow passes through a medium through which the high energy photons of the gamma-ray burst itself have already passed, which is not taken into account in our model.

\section{Implications for GRB afterglow spectra}
   \label{sec-discus}
\begin{figure}
   \centering
   \resizebox{\hsize}{!}{\includegraphics[width=0.95\textwidth,angle=-90]{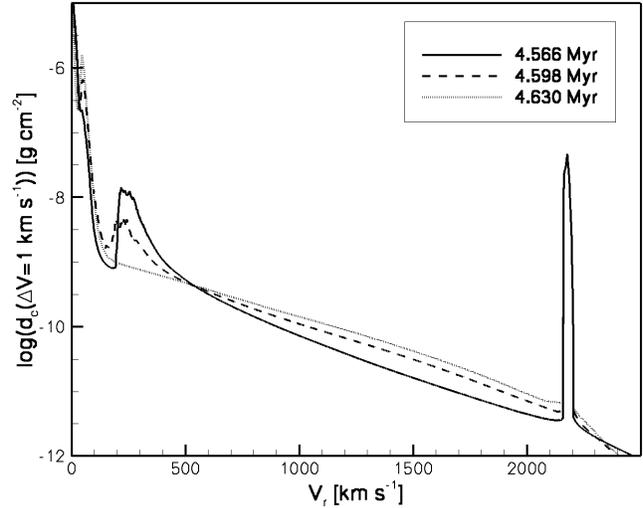}}
      \caption{Angle-averaged column density per velocity interval $\mathrm{d}_{\mathrm{c}}(\varv,\Delta \varv)$ with $\Delta \varv$ = 1 $\kms$ as function of the radial velocity at the same moments in time as the circumstellar medium plots in Figs. \ref{fig:WRbubble} to \ref{fig:WRbubble3}. 
      The formation of the 50 $\kms$ line can be observed, as well as the disappearance of the intermediate velocity feature (cf. also Fig. \ref{fig:cd_all}).
      }
         \label{fig:cd_cuts}
   \end{figure}

     Figures \ref{fig:cd_all} to \ref{fig:stat} show that circumstellar absorption features, to be expected in a gamma-ray burst afterglow spectrum, strongly depend on the duration of the Wolf-Rayet phase of the progenitor star. 
     This may enable us to draw several conclusions regarding a gamma-ray burst progenitor and its surrounding medium. 

    \subsection{The progenitor star} 
     Absorption lines at zero velocity and the Wolf-Rayet wind velocity might be observed throughout the Wolf-Rayet phase.
     However, the absorption line at intermediate velocity is only visible for a short period of time (approx. 80\,000 \mbox{yr}). 
     The exact speed at which this line appears depends on the relative mass-loss rates and wind velocities during the Wolf-Rayet phase and the preceding stage. 
     This line, caused by the Wolf-Rayet wind-driven shell, reaches its final velocity while the shell is expanding into the freely expanding RSG wind. 
     After it hits the RSG shell at the wind termination shock, the high thermal pressure in the hot bubble may slow down the shell fragments, although this does not happen with this particular model.\\

     It is possible to use an analytical estimate of the shell velocity during its earlier phase, in order to find the velocity at which the absorption line will appear. 
     The velocity of an energy-driven shell, expanding into a previously emitted stellar wind, is given by Kwok (\cite{Kwok}) as:
     \begin{equation}
      \label{eq:shellv}
         V_{\mathrm{s}} = \biggl( \frac{ \dot{m}_2 V_2^2 V_1 }{ 3  \dot{m}_1  } \biggr)^{1/3},\\
     \end{equation}
     with $V_{\mathrm{s}}$ the velocity of the shell, $V_{\mathrm{1}}$ and $\dot{m}_1$ the velocity and mass-loss rate of the old wind, and $V_{\mathrm{2}}$ and $\dot{m}_2$ the velocity and mass-loss rate of the new wind. 
     With typical velocities and mass-loss rates for RSG and Wolf-Rayet winds, this yields a shell velocity on the order of 200 $\kms$, which is indeed what we found. 
     However, both wind velocity and mass-loss rate depend on the metallicity and the initial mass of the star, and they can vary even for one star, if the stellar wind is not spherically symmetric. 
     Since rapidly rotating stars have aspherical wind distributions and gamma-ray burst progenitors are thought to be rapidly rotating stars, this is a rational scenario. 
     For heavier stars, which pass through a Luminous Blue Variable stage rather than a red supergiant stage, the values would be different as well (Garc{\a'i}a-Segura et al. \cite{GML96}).
     Whether the line will be visible at all depends ,of course, on the exact moment at which the star collapses. 
     If the line is visible, it would indicate a short Wolf-Rayet phase, i.e. no more than about 50\,000 to 80\,000 years.
 
     It is also possible that the freely expanding wind area around the star is much larger. 
     Possible explanations for this would be a low density interstellar medium, or a lower mass-loss rate during the main sequence. 
     Either would create a lower thermal pressure in the main sequence bubble, which in turn would place the wind termination shock further away from the star. 
     That this can indeed occur would seem to follow from the calculations of Chevalier et al. (\cite{CLF04}), which give a wind termination shock 28 \mbox{pc} from the central star for \object{GRB 990123}.
     The intermediate velocity and 50 $\kms$ lines can be used to find out how much time the star spent as a Wolf-Rayet star before producing a supernova. 

     The intermediate velocity line is only visible during WR1 and the first part of WR2, although the absence of this line may be the result of the position of the shell fragments (see Sect. \ref{sec-coldens}).
     The line at 50 $\kms$ appears later, marking the transition from WR2 to WR3, and thus providing another reference point for the age of the progenitor star, although we cannot be certain of being able to identify this line at all, since it is so close to the zero velocity line.\\

     \subsection{The circumstellar medium}
     While some gamma-ray burst afterglows indicate a $1/r^2$ density distribution in the circumstellar medium, others appear to indicate a constant density (Chevalier et al. \cite{CLF04}).
     The only areas with constant density in the circumstellar medium are the hot bubbles of shocked wind material, which drive the shells outward. 
     If the gamma-ray burst afterglow indicates a constant density medium, this means that the relativistic jet has reached the hot bubble and therefore already passed the freely expanding wind region.
     Consequently,  in a constant density afterglow profile, the high velocity line caused by the Wolf-Rayet wind should not be visible, since the afterglow source is already between that area and the observer. 
     Similarly, the presence of a high velocity absorption line would seem to indicate that the afterglow was caused in a $1/r^2$ density region. \\

     \subsection{GRB~021004}
     \label{sec-GRB021004}
     The best observations so far of the spectrum of a Gamma-ray burst afterglow have been made of \object{GRB 021004}. 
     Identification of the \ion{C}{IV} and \ion{Si}{IV} lines indicate distinctive absorption lines at velocities: 0 $\kms$, 560 $\kms$, and 3000 $\kms$ (Schaefer et al. \cite{Setal03}), although the last line may actually consist of two separate components, both at high velocities.  
     This pattern agrees qualitatively with our column density simulations during the early Wolf-Rayet phase, which would indicate that the progenitor star only spent a short time as a Wolf-Rayet star. 
     A faster Wolf-Rayet wind ($\sim$ 3000 $\kms$ rather than 2200 $\kms$) remains within acceptable parameters for massive stars. 
     Shell velocities of $\sim$ 500 $\kms$ can be reached by adjusting the wind velocities and mass-loss rates of the red supergiant and Wolf-Rayet winds (cf. Eq. \ref{eq:shellv}). 
     Calculations done by Li \& Chevalier (\cite{LC03}) indicate that the afterglow of \object{GRB 021004} results from interaction with stellar wind rather than a constant density medium. 
     This agrees with the presence of high velocity lines in the spectrum.
     If there are indeed two high velocity lines instead of one, this might be explained by a two-stage Wolf-Rayet wind (a fast wind, followed by a slower wind as the star expands due to He-shell burning), or by the presence of clumps in the Wolf-Rayet wind, or by acceleration of part of the Wolf-Rayet wind by radiation from the Gamma-ray burst itself.

\begin{figure}
   \centering
   \resizebox{\hsize}{!}{\includegraphics[width=0.95\textwidth,angle=-90]{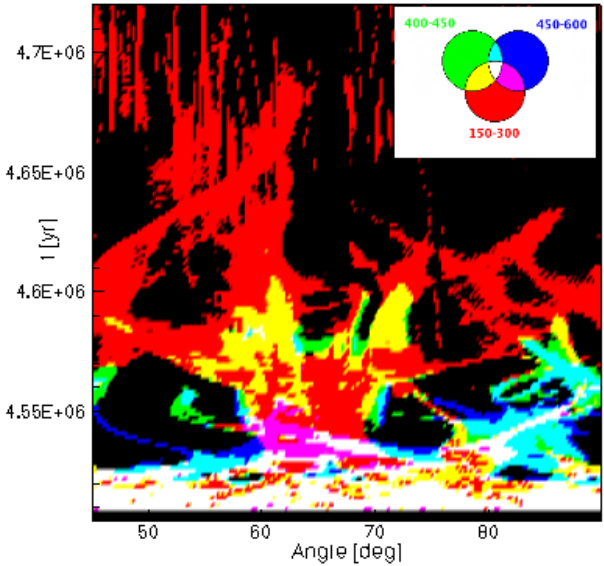}}
      \caption{The presence of gas with a radial column density above a threshold value ($10^{-9}$ \mbox{g} \mbox{cm}$^{-2}$) moving in three velocity intervals, as a function of angle and time. 
      The angle is measured from the pole down, so an angle of 90 degrees is equivalent to the equatorial plane. 
      Red color indicates the presence of gas in the 150-300 $\kms$ interval, while green and blue indicate the same for the 400-450 $\kms$ and 450-600 $\kms$ intervals, respectively. 
      Mixed colors indicate that more than one interval has a high enough column density; e.g., white indicates the presence of gas with high enough column density in all three intervals.
      Pink areas are of particular interest, as these indicate a high column density in both the 150-300 $\kms$ and the 450-600 $\kms$ intervals, whereas the 400-450 $\kms$ interval has a low column density. 
      This means that the intermediate velocity feature will show two well-separated components (see also Fig. \ref{fig:oneangle}).
      }
         \label{fig:angledep}
   \end{figure}

\begin{figure}
   \centering
   \resizebox{\hsize}{!}{\includegraphics[width=0.95\textwidth,angle=-90]{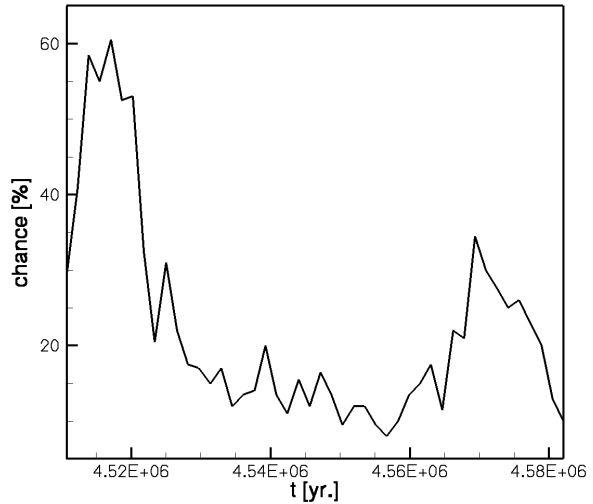}}
      \caption{The probability of observing the intermediate velocity feature as function of time. 
      To estimate this probability, we have taken the 2D results during the early Wolf-Rayet phase (WR1 and the first half of WR2) and checked along each radial gridline whether the intermediate velocity line would be visible. 
      The criterion for this was a five-fold increase in the column density between the area around 100 $\kms$ and the area around 200 $\kms$. 
      }
         \label{fig:stat}
   \end{figure}

\begin{figure}
   \centering
   \resizebox{\hsize}{!}{\includegraphics[width=0.95\textwidth,angle=-90]{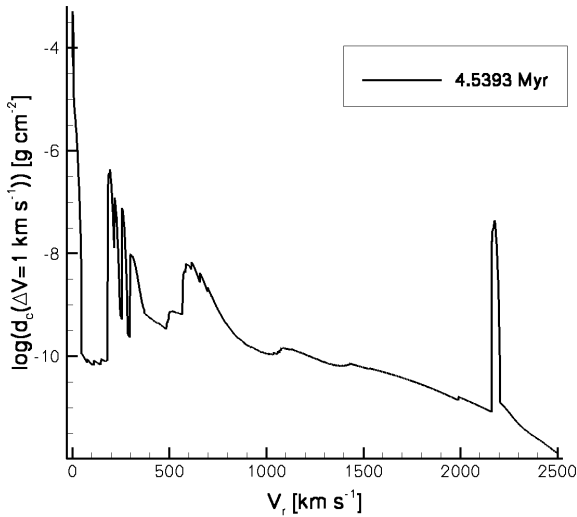}}
      \caption{Column density along a single line of sight, i.e. 
       $\mathrm{d}_{\mathrm{c}}(\varv,\Delta \varv)$ with $\Delta \varv$ = 1 $\kms$ for an angle of 61$^o$ at time t = 4.5393 \mbox{Myr}. 
        Angle and time were chosen from Fig. \ref{fig:angledep} so that two well-separated components of the intermediate velocity feature are present. 
      }
         \label{fig:oneangle}
   \end{figure}

\begin{figure}
   \centering
   \resizebox{\hsize}{!}{\includegraphics[width=0.95\textwidth,angle=-90]{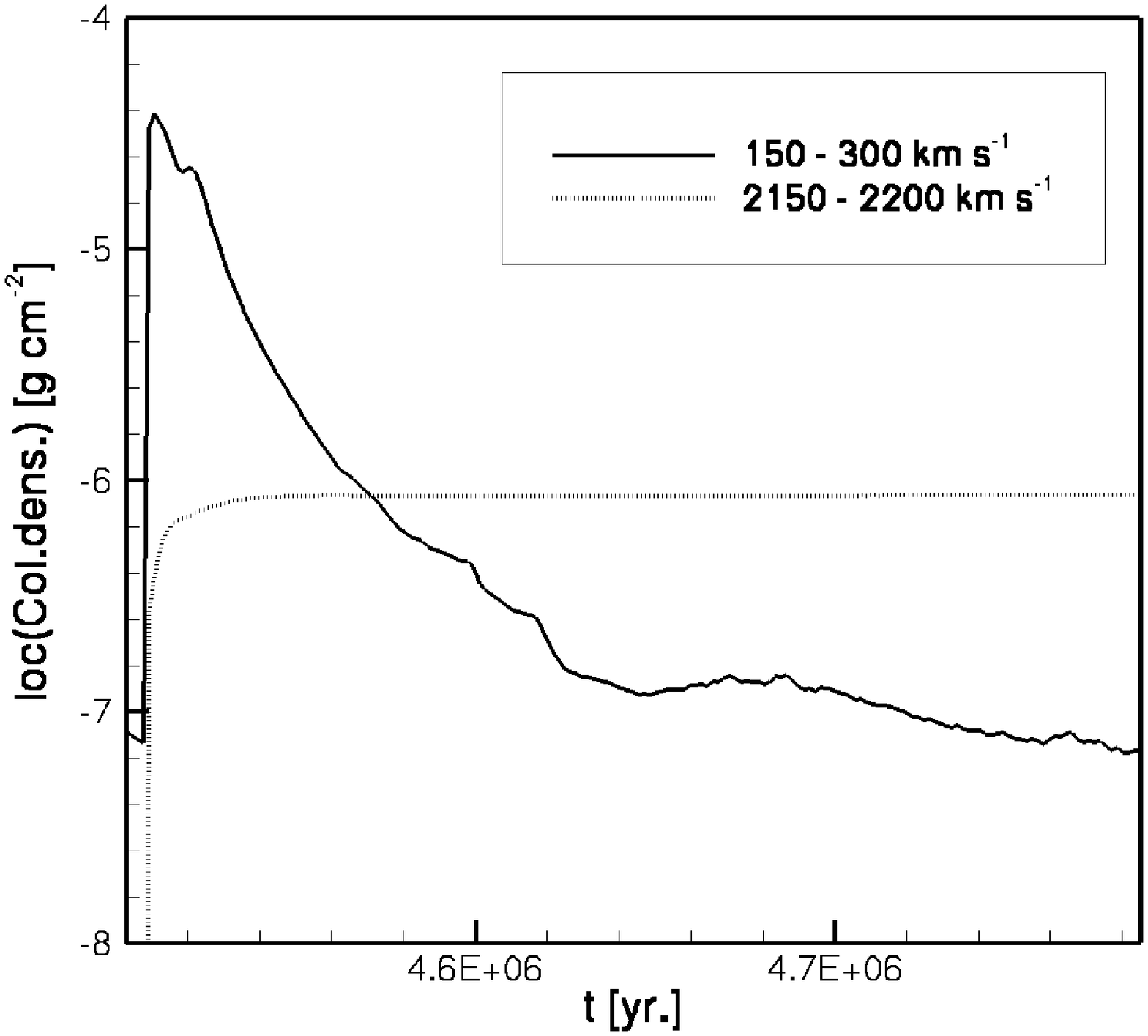}}
      \caption{The total column density of the intermediate velocity and 2200 $\kms$ lines, respectively. 
      The column density of all velocity intervals between 150 and 300 $\kms$ and 2150 and 2200 $\kms$ has been added up. 
      Clearly, once the Wolf-Rayet wind starts, the total column density within the Wolf-Rayet wind remains more or less constant. 
      The intermediate velocity feature is extremely time dependent, since this line is caused by a moving shell that expands into a spherical volume. 
      Note that these column densities where averaged over 200 radial lines. 
      This makes little difference for the Wolf-Rayet wind line, since the density of the wind is barely angle dependent. 
      The column density in the intermediate velocity line, however, is angle dependent, as is shown in Fig. \ref{fig:angledep}.
      }
         \label{fig:linecoldens}
   \end{figure}

     VLT spectra of \object{GRB 021004} were presented by Fiore et al. (\cite{Fietal05}).
     These observations show a larger number of absorption lines at low velocities, leading to a total of six different lines in both \ion{C}{IV} and \ion{Si}{IV}. 
     The large number of intermediate velocity components (3) can be explained as the result of multiple clumps of the shell fragments appearing in the line of sight (see Sect. \ref{sec-coldens}).
     The column densities for all these lines are on the order of $10^{14}$ to $10^{15}$ particles per \mbox{cm}$^2$. 
     In order to compare these to our own results, we calculated the total column densities in the intermediate velocity and 2200 $\kms$ lines. 
     The results in Fig. \ref{fig:linecoldens} show that total column densities for the Wolf-Rayet wind line are nearly constant throughout the Wolf-Rayet phase. 
     Not surprisingly so, since the mass loss rate and wind velocity remain unchanged. 
     The location of the wind termination shock varies, so the total Wolf-Rayet wind area changes over time, but this takes place at a range where the wind density is extremely low (see Figs. \ref{fig:WRbubble} to \ref{fig:WRbubble3}), therefore the change in column density is minimal. 
     The column density of the intermediate velocity feature is strongly time dependent. 
     After all, this feature is the result of a moving shell which expands into a spherical volume;
     therefore, the density in the shell will decrease with $\rho \sim \mathrm{r}^{-2}$ until the shell remnants hit the outer edge of the bubble. 
     Column densities for the lines in Fig.\ref{fig:linecoldens} are on th eorder of $10^{-5}$ to $10^{-7}$ in \mbox{g} \mbox{cm}$^{-2}$, which corresponds to $10^{19}$ to $10^{16}$ \mbox{amu} per \mbox{cm}$^{-2}$. 
     As carbon has an average weight of about 12 \mbox{amu}, this amounts to a particle column density of $10^{18}$ to $10^{15}$ carbon ions per \mbox{cm}$^{-2}$, if the gas were pure carbon. 
     Obviously, this is not the case, so the column density of carbon ions would probably be on the order of 0.1 to 10 percent, which corresponds quite well with the figures found by Fiore et al. (\cite{Fietal05}).
     We should keep in mind that in Fig. \ref{fig:linecoldens} the column densities have been averaged over the 200 radial lines. 
     This does not affect the Wolf-Rayet wind line, since the density of the wind is more or less spherically symmetric. 
     However, the shell fragments, which make up the intermediate velocity line do not have such a symmetry, so column densities in the line of sight may vary considerably (see Figs. \ref{fig:angledep} through \ref{fig:oneangle}).     

     A pattern of blue-shifted absorption lines similar to those observed in a gamma-ray burst afterglow may be visible in the spectra of Type Ib/c supernovae, since these supernovae also have Wolf-Rayet stars as progenitors. 
     Dopita et al. (\cite{DECS84}) attributed narrow P Cygni profiles seen in H$\mathrm{\alpha}$ and H$\mathrm{\delta}$ in the spectrum of \object{SN 1984E} to a Wolf-Rayet wind with a velocity of about 3000 $\kms$. 
     Similarly, Bowen at al. (\cite{BRMB00}) and Fassia et al. (\cite{Fetal01}) note the presence of a broad absorption feature covering $\sim$ 350 $\kms$ in the spectrum of \object{SN 1998S}, which may be caused by a moving shell.    

\section{Final remarks} 
\label{sec-finrem}    
 In this paper, we describe a powerful method for constraining the progenitors of GRBs and SNe through narrow circumstellar spectral lines (see also van Marle et al. \cite{MLG05}).  
 While we expect a Type Ib/c supernova from practically all massive stars whose hydrogen-rich envelope is stripped, only a small fraction of these stars --- 
i.e. $(0.002 \dots 0.004) $ (van Putten \cite{P04}) --- produces a gamma-ray burst. 
 We have shown that the characteristics of the circumstellar line spectrum change significantly during the Wolf-Rayet evolution --- in line with the fact that only a fraction of the Galactic Wolf-Rayet stars are still surrounded by a nebula --- which means that the circumstellar lines which are observable in a GRB afterglow or in a supernova explosion produced by a Wolf-Rayet star are sensitive indicators of the duration of the preceding Wolf-Rayet phase.

 According to stellar evolution models, the duration of the Wolf-Rayet phase, which is the final evolutionary phase before its death, can vary between almost zero and $10^6\,$yr, where it is larger for higher mass and/or metallicity (Meynet \& Maeder \cite{MM05}).
 The fact that only part of the Galactic Wolf-Rayet stars actually show a nebula means that those tend to disperse on a time scale which is shorter than the mean life time of Wolf-Rayet stars, about $5 \times 10^5$yr.  
 Thus,  Wolf-Rayet nebulae evolve on the evolutionary time scale of their central star.
 Therefore, the corresponding absorption lines at the time the star explodes may be used, e.g., to distinguish whether a Wolf-Rayet star died young or old.

In principle, the evolution of main sequence and Wolf-Rayet bubbles are 
influenced --- although weakly --- 
by various stellar and interstellar medium parameters (Weaver et al. \cite{WEA77}). 
However, the life 
time of the Wolf-Rayet shell, or of the intermediate velocity absorption
line feature, is on the order of shell radius divided by shell velocity. 
For shell velocities obtained
within our model, this time scale is some 10\,000\,yr. As the shell velocities
coincide well with the values derived for the intermediate velocity absorption 
components in the afterglow spectra of \object{GRB 021004}, the conclusion that
the Wolf-Rayet phase of its progenitor lasted a much shorter time than the average Wolf-Rayet
life time is in fact independent of the particular choice of stellar and 
interstellar medium parameters used in our model.

A short Wolf-Rayet phase of the progenitor of \object{GRB 021004}
rules out that the underlying Wolf-Rayet star was formed through the LBV-scenario (cf. Garc{\a'i}a-Segura et al. \cite{GML96}), since in this case the duration of the Wolf-Rayet phase corresponds to the full duration of core helium burning (about $5 \times 10^5$yr), and the nebula would be expected to have dispersed by the time the central star dies.
 Early mass transfer (so called Case~A or Case~B; cf. Wellstein \& Langer \cite{WL99}), which --- just like LBV type mass loss --- already produces an early Wolf-Rayet phase during core helium burning, can also be ruled out.

 The presence of an intermediate velocity component in the afterglow spectrum of \object{GRB 021004} indicates that the Wolf-Rayet shell was still intact when the star exploded, i.e. that the Wolf-Rayet phase was relatively short. 
 This could imply a progenitor zero age main sequence (ZAMS) mass close to the lower ZAMS mass limit for Wolf-Rayet star formation, which is at about 25$\mso$ for solar metallicity (Meynet \& Maeder \cite{MM05}) and larger for smaller~$Z$.
 Alternatively, the progenitor of \object{GRB 021004} could have evolved into the Wolf-Rayet phase through so-called Case~C mass transfer, which occurs at the end of core helium burning, and which is assumed to give rise to a common envelope phase (cf. Podsiadlowski et al. \cite{PJH92}).
 In that case, the ZAMS mass of the progenitor would have to be smaller than the lower ZAMS mass limit for Wolf-Rayet star formation.
 
 A short duration of the Wolf-Rayet phase might also avoid strong angular momentum loss of the stellar core associated with the strong Wolf-Rayet mass loss (Langer \cite{L98},
Petrovic et al. \cite{PE05}), which otherwise might reduce the core angular momentum below the critical value required within the collapsar model (MacFadyen \& Woosley \cite{MW99}).
 A late common envelope phase, on the other hand, might even lead to a spin-up of the Wolf-Rayet star, either through tidal interaction in a tight post-common envelope binary system (e.g. van Putten \cite{P04}) or through angular momentum gain in the course of a  merger event (e.g., Fryer \& Heger \cite{FH05}). 

 Observed circumstellar spectral lines for a larger number of  SNe~Ib/c or GRB afterglows might enable us to empirically constrain the subsample of Wolf-Rayet stars that are capable of producing a gamma-ray burst. 
 Since prototype observations exist for both cases, we may be able to reach this aim within a certain period of time. 
 However, more models of the circumstellar medium of massive supernova progenitors at various metallicities, and for single stars and binary systems, are also needed before  more general conclusion can be drawn.

\begin{acknowledgements} 
 We would like to thank Alexander van der Horst, Rhaana Starling, Klaas Wiersema and Ralph Wijers for their information on the spectrum of the afterglow of \object{GRB 021004}.
 This work was sponsored by the Stichting Nationale Computerfaciliteiten (NCF), with financial support from the Nederlandse Organisatie voor Wetenschappelijk Onderzoek (NWO).
 This research was done as part of the AstroHydro3D project:~(http://www.strw.leidenuniv.nl/AstroHydro3D/)
\end{acknowledgements}

\listofobjects

\end{document}